\documentstyle[aps]{revtex}

\begin{document}
\title{Symmetric Hyperbolic System in the Self-dual Teleparallel Gravity}
\author{G. Y. Chee}
\address{Physics Department, Liaoning Normal University, Dalian, 116029, China}
\author{Yongxin Guo}
\address{Physics Department, Liaoning University, Shenyang, 110036, China}
\maketitle

\begin{abstract}
In order to discuss the well-posed initial value formulation of the
teleparallel gravity and apply it to numerical relativity a symmetric
hyperbolic system in the self-dual teleparallel gravity which is equivalent
to the Ashtekar formulation is posed. This system is different from the ones
in other works by that the reality condition of the spatial metric is
included in the symmetric hyperbolicity and then is no longer an independent
condition. In addition the constraint equations of this system are rather
simpler than the ones in other works.

\end{abstract}

\section{Introduction}

As the closest alternative to general relativity (GR), teleparallel gravity
can be traced back to Einstein [1] who regarded it as a unified field theory
and attempted to use it to supersede GR. Teleparallel gravity can be
regarded as a translational gauge theory [2-4], which makes it possible to
unify gravity with other kinds of interactions in the gauge theory
framework. Poincare gauge theory is a natural extension of the gauge
principle to spacetime symmetry , and represents a alternative to GR (for
more general attempts see [3]).In particular, teleparallel gravity was
regarded as a promising alternative to GR until the work of Kopczynski [5],
who found a hidden gauge symmetry which prevents the torsion from being
completely determined by the field equations, and concluded that the theory
is inconsistent. Nester [6] improved the arguments by showing that the
unpredictable behavior of torsion occurs only for some very special
solutions (see also [7]). Hecht et al. [8] investigated the initial value
problem of teleparallel gravity and conclude that it is well defined if the
undetermined velocity are dropped out from the set of dynamic velocities.

Teleparallel gravity possesses many salient features. Because of its
simplicity and transparency teleparallel gravity seems to be much more
appropriate than GR to deal with the problem of gravitational energy
momentum [4, 9, 10]. Nester [9] succeeded in proving the positivity of total
energy for Einstein's theory in terms of teleparallel geometry. He found
that special gauge features of teleparallel gravity, which are usually
considered to be problematic, are quite beneficial for this purpose. Mielke
[10] used the teleparallel approach to give a transparent description of
Ashtekar's new variables[11]. Andrade et al. [12] formulated a
five-dimensional equivalent of Kaluza-Klein theory.  Although quantum
properties of the Poincare gauge theory are, in general, not so attractive,
the related behavior in the specific case of the teleparallel theory might
be better [13], and should be further explored.

The canonical Hamiltonian approach is the best way to clarify both the
nature of somewhat mysterious extra gauge symmetries and the question of
consistency of teleparallel gravity. It is found [7] that the presence of
nondynamical torsion components is not a sign of an inconsistency, but a
consequence of the constraint structure of the theory. In some works[7,14,
15], the Hamiltonian formulation of teleparallel gravity has been developed.
However, the well-posed initial value formulation of teleparallel gravity
has not been discussed as yet. As is well known, hyperbolic formulation of
the Einstein equation is one of the main research areas in GR [16]. This
formulation is used in the proof of the existence, uniqueness, and stability
of the solutions of the Einstein equation by analytical methods. Thus far,
several first-order hyperbolic formulations are proposed [17-20]. The recent
interest in hyperbolic formulation arises from its application to numerical
relativity [21]. It is proved that the Einstein equation in Ashtekar's
variables constitutes a symmetric hyperbolic system [22, 23]. A question
naturally arises whether there is a well-posed initial value formulation of
teleparallel gravity which is equivalent to or different from the hyperbolic
formulation of the Einstein equation. If it exists, can it give us some new
perspectives and be applied to numerical relativity? An answer to this
question will be given in this paper. A self-dual teleparallel formulation
of general relativity which is equivalent to Ashtekar's formulation has been
developed, its canonical Hamiltonian analysis has been given and used to
clarify the gauge structure of the theory [15]. In this paper a new
symmetrical hyperbolic system of the Einstein equation will be posed in
terms of the two-spinor formulation based on [15]. A new fact we find is
that the reality condition of the spatial metric is included in the
symmetric hyperbolicity and then is not independent. In addition, the
constraint conditions take a rather simple form. All of these reduces the
number of independent conditions imposed on the equations and then
simplifies the relevant problems largely. In Sec. II, a canonical
formulation of the self-dual teleparallel equivalent of GR is given, the
Hamiltonian (evolution) equations and the constraint equations are written
in terms of the two-spinor dyad and the canonical conjugate momenta. In Sec.
III, by introducing a new variable the evolution equations are rewritten as
first-order forms. Then in Sec. IV, the conditions of the symmetric
hyperbolicity of the evolution equations are discussed and the relations
between the conditions of the symmetric hyperbolicity and the reality
conditions of the spatial metric are obtained. Section V is devoted to some
conclusions..

\section{The canonical formulation of the self-dual teleparallel equivalent
of GR}

We start with the Lagragian of the self-dual (or chiral) teleparallel
formulation of general relativity [15,10] which is equivalent to the
Ashitekar Lagrangian [11] and is written in terms of the two-spinor
formulation [24] 
\begin{equation}
V_{||}^{+}=N\sigma [\omega _{(AB)}{}^{AC}\omega ^{DB}{}_{DC}-\omega
_{(AB)CD}\omega ^{CBAD}-\sqrt{2}\omega _{\perp CD}\omega ^{(CE)}{}_E{}^D],
\end{equation}
where $\omega _{AB}{}^{CD}$ is the self-dual spin connection, $\omega
_{(AB)}{}^{CD}$ is the Ashtekar variable, 
\begin{equation}
\omega _{\perp CD}=n^{AB}\omega _{ABCD}=n^{AB}\omega _{[AB]CD},
\end{equation}
and 
\begin{equation}
n^{AB}=\frac 1N(t^{AB}-N^{AB}),
\end{equation}
with the determinant of the inverse soldering form $\sigma =\det \sigma _\mu
{}^{AB}=\frac 1N\sqrt{-g}$ , the lapse $N$ and the shift $N^{AB}$.

In the two-spinor formalism [15] the basic variable is chosen to be the dyad 
$\zeta _{aA}$: 
\[
\zeta _{0A}=o_A,\zeta _{1A}=\iota _A,
\]
and the self-dual spin connection $\omega _{ABCD}$ can be expressed by 
\begin{equation}
\omega _{ABCD}=\zeta _C{}^b\partial _{AB}\zeta _{bD}.
\end{equation}
Using (2) (3) and (4) one gets 
\begin{equation}
\omega _{\perp CD}=-\frac 1N\zeta _C{}^b\stackrel{\cdot }{\zeta }_{bD}-\frac %
1NN^{AB}\omega _{(AB)CD},
\end{equation}
where 
\[
\stackrel{\cdot }{\zeta }_{bD}=t^{AB}\partial _{AB}\zeta _{bD}.
\]
Then the Lagrangian $V_{||}^{+}$ becomes 
\begin{eqnarray*}
V_{||}^{+} &=&N\sigma [\omega _{(AB)}{}^{AC}\omega ^{(DB)}{}_{DC}-\omega
_{(AB)CD}\omega ^{(CB)AD}] \\
&&-2\sqrt{2}\sigma \zeta _C{}^b\stackrel{\cdot }{\zeta }_{bD}\omega
^{(EC)}{}_E{}^D+2\sqrt{2}\sigma N^{AB}\omega _{(AB)CD}\omega ^{(EC)}{}_E{}^D.
\end{eqnarray*}
The canonical momentum conjugate to $\zeta _{bB}$ is then 
\begin{equation}
\widetilde{p}^{bD}=\frac{\partial V_{||}^{+}}{\partial \stackrel{\cdot }{%
\zeta }_{bD}}=-2\sqrt{2}\sigma \zeta _C{}^b\omega ^{(EC)}{}_E{}^D,
\end{equation}
which leads to 
\[
\omega _{(AC)}{}^{CB}=\frac{\sqrt{2}}{4\sigma }\zeta _{aA}{}{}\widetilde{p}%
^{aB}.
\]
The gravitational Hamiltonian can be computed 
\begin{equation}
{\cal H}_G^{+}=\widetilde{p}^{bD}\stackrel{\cdot }{\zeta _{bD}}-V_{||}^{+}=N%
{\cal H}_{\perp }+N^{AB}{\cal H}_{AB},
\end{equation}
where 
\begin{equation}
{\cal H}_{\perp }=\frac 1{8\sigma }\epsilon _{ab}{}{}\widetilde{p}^{aC}%
\widetilde{p}^b{}_C+\sigma \zeta ^a{}_C\zeta ^{bA}\partial _{(AB)}\zeta
_{aD}\partial ^{(CB)}\zeta _b{}^D,
\end{equation}
and 
\begin{equation}
{\cal H}_{AB}=\partial _{(AB)}\zeta _{aD}\widetilde{p}^{aD}.
\end{equation}
Computing the variation, we obtain the Hamiltonian equations 
\begin{eqnarray}
\stackrel{\cdot }{\zeta }_{aA} &=&\frac{\delta {\cal H}_G}{\delta \widetilde{%
p}^{aA}}=-\frac N{4\sigma }{}{}\widetilde{p}_{aA}+N^{CB}\partial
_{(CB)}\zeta _{aA}, \\
\stackrel{\cdot }{\widetilde{p}^{aA}} &=&-\frac{\delta {\cal H}_G^{+}}{%
\delta \zeta _{aA}}  \nonumber \\
&=&N^{CB}\partial _{(BC)}\widetilde{p}^{aA}+2N\sigma \zeta ^a{}_D\zeta
^{bB}{}\partial _{(BC)}\partial ^{(DC)}\zeta _b{}^A-\frac N{2\sigma }\zeta
^{aA}{}{}\widetilde{p}^{bB}\widetilde{p}_{bB}  \nonumber \\
&&-4N\sigma \zeta ^{aA}\zeta ^c{}_C\zeta ^{bE}\partial _{(EB)}\zeta
_{cD}\partial ^{(CB)}\zeta _b{}^D+8N\sigma \zeta ^a{}_D\zeta ^{bB}\zeta
^{cE}\partial _{(BC)}\zeta _{cE}{}\partial ^{(DC)}\zeta _b{}^A  \nonumber \\
&&-2N\sigma \zeta ^{bC}\partial _{(CB)}\zeta ^{aD}{}\partial ^{(AB)}\zeta
_{bD}{}+2N\sigma \zeta ^{bB}\partial _{(BC)}\zeta ^a{}_D{}\partial
^{(DC)}\zeta _b{}^A  \nonumber \\
&&+2N\sigma \zeta ^a{}_D\partial _{(BC)}\zeta ^{bB}{}\partial ^{(DC)}\zeta
_b{}^A+\partial _{(BC)}N^{CB}\widetilde{p}^{aA}-2\partial _{(BC)}N\sigma
\zeta ^a{}_D\zeta ^{bB}{}\partial ^{(DC)}\zeta _b{}^A,
\end{eqnarray}
and the constraint equations 
\begin{equation}
{\cal H}_{\perp }=0,{\cal H}_{AB}=0.
\end{equation}
The detail of the constraint structure of the theory can be found in [15].

\section{The first-order evolution equations}

Since the Hamiltonian equations of $\widetilde{p}^{aA}$ include the
second-order terms $2N\sigma \zeta ^a{}_D\zeta ^{bB}{}\partial
_{(BC)}\partial ^{(DC)}\zeta _b{}^A$, in order to get the first-order
evolution equations we decompose $\omega _{(AC)}{}^{DB}$ into its trace-free
and trace part 
\begin{equation}
\omega _{(AC)}{}^{DB}=\omega _{trf(AC)}{}^{DB}+\frac 1{4\sqrt{2}\sigma }%
\epsilon _C{}^D\zeta _{aA}{}{}\widetilde{p}^{aB}
\end{equation}
and introduce a new variable $\widetilde{q}_{AB}$ and a real constant
spatial 1-form $\psi _{CD}$ by 
\begin{equation}
\omega _{trf(AC)DB}=\frac{\sqrt{2}}{8\sigma }\psi _{D(C}\widetilde{q}_{A)B},
\end{equation}
with the properties 
\begin{equation}
\psi _{CD}=\psi _{(CD)},\;\widetilde{q}_{AB}=\widetilde{q}_{[AB]}.
\end{equation}
Substituting (14) into (13) and using (4)\ we obtain

\begin{equation}
\partial _{(AC)}\zeta _{aB}=\frac{\sqrt{2}}{8\sigma }[\psi _{a(C}\widetilde{q%
}_{A)B}{}-\zeta _{a(C}\widetilde{p}_{A)B}{}].
\end{equation}
Then we can compute 
\begin{equation}
\stackrel{\cdot }{\zeta }_{aA}=-\frac N{4\sigma }{}{}\widetilde{p}_{aA}-%
\frac{\sqrt{2}N^{CB}}{8\sigma }\zeta _{aB}\widetilde{p}_{CA}+\frac{\sqrt{2}%
N^{CB}}{8\sigma }\psi _{aB}\widetilde{q}_{CA}{}{},
\end{equation}
and 
\begin{equation}
\stackrel{\cdot }{\sigma }=-[N{}{}\zeta ^{aA}\widetilde{p}_{aA}+\frac{\sqrt{2%
}N^{CB}}2\psi ^A{}_B\widetilde{q}_{CA}{}].
\end{equation}

Using these results we can rewrite (11) as 
\begin{eqnarray}
\stackrel{\cdot }{\widetilde{p}^{aA}} &=&\frac{\sqrt{2}N}8\zeta ^{aB}{}\zeta
_b{}^C\partial _{(BC)}\widetilde{p}^{bA}+N^{CB}\partial _{(BC)}\widetilde{p}%
^{aA}  \nonumber \\
&&-\frac{\sqrt{2}N}8\zeta ^a{}_D\psi {}^{BC}\partial _{(BC)}\widetilde{q}%
^{DA}+\frac{\sqrt{2}N}8\psi {}^{aB}\partial _{(BC)}\widetilde{q}^{CA} 
\nonumber \\
&&+Q_p(p,q),
\end{eqnarray}
where source term $Q_p(p,q)$ is a quadratic polynomial of $\widetilde{p}%
^{aA} $ and $\widetilde{q}_{AB}$.

Introducing a constant spatial vector $\varphi ^{AB}$ satisfying 
\begin{equation}
\psi _{AC}\varphi ^{AB}=\epsilon _C{}^B
\end{equation}
and using (16) one can obtain 
\begin{equation}
\widetilde{q}_{AB}=\frac{8\sqrt{2}}3\sigma \zeta {}^{bD}\varphi
_D{}^C\partial _{(AC)}\zeta _{bB}-\frac 13\varphi _{AC}{}\zeta _a{}{}{}^C%
\widetilde{p}^a{}_B.
\end{equation}
Taking the time derivative leads to 
\begin{eqnarray}
\stackrel{\cdot }{\widetilde{q}}_{AB} &=&\frac{8\sqrt{2}}3\sigma \zeta
{}^{bD}\varphi _D{}^C\partial _{(AC)}\stackrel{\cdot }{\zeta }_{bB}-\frac 13%
\varphi _{AC}{}\zeta _a{}{}{}^C\stackrel{\cdot }{\widetilde{p}}^a{}_B 
\nonumber \\
&&+\frac{8\sqrt{2}}3\stackrel{\cdot }{\sigma }\zeta {}^{bD}\varphi
_D{}^C\partial _{(AC)}\zeta _{bB}+\frac{8\sqrt{2}}3\sigma \stackrel{\cdot }{%
\zeta }{}^{bD}\varphi _D{}^C\partial _{(AC)}\zeta _{bB}  \nonumber \\
&&-\frac 13\varphi _{AC}{}\stackrel{\cdot }{\zeta }_a{}{}{}^C\widetilde{p}%
^a{}_B.
\end{eqnarray}
Using (17), (18), and (21) we get the evolution equation of $\widetilde{q}%
_{AB}$: 
\begin{eqnarray}
&&\stackrel{\cdot }{\widetilde{q}}_{AB}  \nonumber \\
&=&-\frac{2\sqrt{2}N}3\zeta {}^{aD}\varphi _D{}^C\partial _{(AC)}{}{}%
\widetilde{p}_{aB}-\frac{\sqrt{2}N}{24}\zeta {}^{aD}\varphi
_A{}{}^C{}\partial _{(CD)}\widetilde{p}{}_{aB}+\frac 23N^{DE}\varphi
_E{}^C\zeta ^a{}_D\partial _{(AC)}\widetilde{p}_{aB}+\frac 13N^{CE}\varphi
_{AD}{}\zeta {}{}{}^{aD}\partial _{(CE)}\widetilde{p}{}_{aB}  \nonumber \\
&&-\frac{\sqrt{2}N}{24}\varphi _{AC}{}\psi {}^{DE}\partial _{(DE)}\widetilde{%
q}^C{}_B-\frac{\sqrt{2}N}{24}\psi {}^{CD}\varphi _{AC}{}\partial _{(DE)}%
\widetilde{q}^E{}_B+\frac 23N^{CD}\partial _{(AC)}\widetilde{q}_{DB} 
\nonumber \\
&&+Q_q(p,q),
\end{eqnarray}
where source term $Q_q(p,q)$ is another quadratic polynomial of $\widetilde{p%
}^{aA}$ and $\widetilde{q}_{AB}$. The source terms $Q_p(p,q)$ and $Q_q(p,q)$
do not contain any derivatives of the fundamental variables other than the
lapse $N$ and the shift $N^{AB}$ .

Substituting (16) into (8) and (9) one can obtain: 
\[
{\cal H}_{AB}=\frac{\sqrt{2}}{16\sigma }[\psi _{CA}\widetilde{q}_{BD}{}+\psi
_{CB}\widetilde{q}_{AD}{}]\widetilde{p}^{CD}, 
\]
and 
\[
{\cal H}_{\perp }=-\frac{13}{128\sigma }\widetilde{p}_{AB}\widetilde{p}%
^{AB}{}{}-\frac 1{64\sigma }\psi ^A{}_B\widetilde{p}_{AC}\widetilde{q}^{BC}-%
\frac{25}{384\sigma }\widetilde{q}_{AB}{}\widetilde{q}^{AB}. 
\]
And then the constraint equations (12) leads to 
\begin{equation}
\widetilde{q}_{AB}{}{}=-\widetilde{q}_{BA},
\end{equation}
and 
\begin{equation}
39\widetilde{p}_{AB}\widetilde{p}^{AB}+6\psi ^A{}_B\widetilde{p}_{AC}%
\widetilde{q}^{BC}+25\widetilde{q}_{AB}{}\widetilde{q}^{AB}=0
\end{equation}
The equations (16), (24) and (25) constitute the constraint equations.
However, the equation (24) just confirms the equation (15) and then is not a
independent constraint. The independent constraints are only (16) and (25).

\section{The symmetric hyperbolicity of the evolution equations}

The principal parts of the evolution equations (17), (19) and (23) are,
respectively 
\begin{equation}
\stackrel{\cdot }{\zeta }_{aA}\cong 0,
\end{equation}

\begin{eqnarray}
&&\stackrel{\cdot }{\widetilde{p}^{aA}}\cong \frac{\sqrt{2}}8N\zeta
^{aB}{}\zeta _b{}^C\partial _{(BC)}\widetilde{p}^{bA}+N^{CB}\partial _{(BC)}%
\widetilde{p}^{aA}  \nonumber \\
&&-\frac{\sqrt{2}}8N\zeta ^a{}_D\psi {}^{BC}\partial _{(BC)}\widetilde{q}%
^{DA}+\frac{\sqrt{2}}8N\psi {}^{Ba}\partial _{(BC)}\widetilde{q}^{CA} 
\nonumber \\
&=&\Phi ^{aADC}{}_{nN}\partial _{(DC)}\widetilde{p}^{nN}+\Psi
^{aADC}{}_{MN}\partial _{(DC)}\widetilde{q}^{MN},
\end{eqnarray}
and 
\begin{eqnarray}
\stackrel{\cdot }{\widetilde{q}}_{AB} &\cong &-\frac{2\sqrt{2}N}3\varphi
^{aC}\partial _{(AC)}{}{}\widetilde{p}_{aB}-\frac{\sqrt{2}}{24}N\varphi
_A{}^D{}\zeta {}^{aC}\partial _{(DC)}\widetilde{p}{}_{aB}+\frac{2N^{DE}}3%
\varphi _D{}^C\zeta ^a{}_E\partial _{(AC)}\widetilde{p}_{aB}+\frac 13%
N^{CD}\varphi _A{}^a\partial _{(DC)}\widetilde{p}{}_{aB}  \nonumber \\
&&-\frac{\sqrt{2}}{24}N\partial _{(AC)}\widetilde{q}^C{}_B{}-\frac{\sqrt{2}}{%
24}N\varphi _{CA}\psi {}^{ED}\partial _{(ED)}\widetilde{q}^C{}_B+\frac{%
2N^{DC}}3\partial _{(AD)}\widetilde{q}_{CB}  \nonumber \\
&=&\Theta _{AB}{}^{CDnN}{}\partial _{(CD)}\widetilde{p}_{nN}+\Pi
_{AB}{}^{CDMN}\partial _{(CD)}\widetilde{q}{}{}_{MN},
\end{eqnarray}
where 
\begin{equation}
\Phi ^{aADC}{}_{nN}=\frac{\sqrt{2}}8N\zeta ^{aD}{}\zeta _n{}^C\epsilon
_N{}^A+N^{CD}\epsilon _n{}^a\epsilon _N{}^A,
\end{equation}
\begin{equation}
\Psi ^{aADC}{}_{MN}=-\frac{\sqrt{2}}8N\zeta ^a{}_M\psi {}^{DC}\epsilon
_N{}^A+\frac{\sqrt{2}}8N\psi {}^{Da}\epsilon _M{}^C\epsilon _N{}^A,
\end{equation}
\begin{equation}
\Theta _{AB}{}^{CDnN}=-\frac{2\sqrt{2}N}3\varphi ^{nD}\epsilon
_A{}^C\epsilon _B{}^N-\frac{\sqrt{2}}{24}N\varphi _A{}^D{}\zeta
{}^{nC}\epsilon _B{}^N+\frac{2N^{FE}}3\varphi _F{}^D\epsilon _A{}^C\zeta
^n{}_E\epsilon _B{}^N+\frac 13N^{CD}\varphi _A{}^n\epsilon _B{}^N,
\end{equation}
and 
\begin{equation}
\Pi _{AB}{}^{CDMN}=-\frac{\sqrt{2}}{24}N\epsilon _A{}^C\epsilon
^{DM}\epsilon _B{}^N{}+\frac{\sqrt{2}}{24}N\varphi ^M{}_A\psi
{}^{CD}\epsilon _B{}^N{}+\frac{2N^{DM}}3\epsilon _A{}^C{}\epsilon _B{}^N.
\end{equation}
The principal parts of the evolution equations can be expressed as a
''matrix form'' 
\begin{equation}
\partial _t\left( 
\begin{array}{l}
\zeta _{aA} \\ 
\widetilde{p}^{aA} \\ 
\widetilde{q}_{AB}{}
\end{array}
\right) =\left( 
\begin{array}{lll}
0 & 0 & 0 \\ 
0 & \Phi ^{aACD}{}_{nN} & \Psi ^{aACDMN}{}{} \\ 
0 & \Theta _{AB}{}^{CD}{}_{nN} & \Pi _{AB}{}^{CDMN}
\end{array}
\right) \partial _{(CD)}\left( 
\begin{array}{l}
\zeta _{bB} \\ 
\widetilde{p}^{nN} \\ 
\widetilde{q}_{MN}{}
\end{array}
\right) ,
\end{equation}
and then they are symmetric hyperbolic if the conditions 
\begin{eqnarray*}
\Phi ^{aACDnN} &=&\overline{\Phi }^{nNCDaA}, \\
\Theta ^{ABCDnN} &=&\overline{\Psi }^{ABCDnN}, \\
\Pi ^{ABCDMN} &=&\overline{\Pi }^{MNCDAB},
\end{eqnarray*}
are satisfied. Using the formulas 
\[
\epsilon {}^{NA}=\epsilon {}^{mb}\zeta _m{}^N\zeta _b{}^A,\;\overline{%
\epsilon {}}^{NA}=\overline{\epsilon }{}^{mb}\overline{\zeta }_m{}^N%
\overline{\zeta }_b{}^A 
\]
and 
\[
\overline{\Phi }^{nNCDaA}=\frac{\sqrt{2}}8\overline{N}\overline{\zeta }%
^{nD}{}\overline{\zeta }{}^{aC}\overline{\epsilon }{}^{AN}+\overline{N}^{CD}%
\overline{\epsilon }{}^{an}\overline{\epsilon {}}^{AN} 
\]
we can find easily that the condition 
\[
\Phi ^{aACDnN}=\overline{\Phi }^{nNCDaA} 
\]
leads to 
\begin{equation}
\frac{\sqrt{2}}8N\zeta ^{aD}{}\zeta {}^{nC}\epsilon {}^{NA}+N^{CD}\epsilon
{}^{na}\epsilon {}^{NA}=\frac{\sqrt{2}}8\overline{N}\overline{\zeta }^{nD}{}%
\overline{\zeta }{}^{aC}\overline{\epsilon }{}^{AN}+\overline{N}^{CD}%
\overline{\epsilon }{}^{an}\overline{\epsilon {}}^{AN}.
\end{equation}
Supposing 
\begin{equation}
\overline{N}^{CD}=N^{CD},
\end{equation}
and using 
\[
\epsilon {}^{mb}=\overline{\epsilon }{}^{mb}, 
\]
we find from (34) that 
\begin{equation}
\epsilon {}^{NA}=\overline{\epsilon }{}^{NA},N=\overline{N}.
\end{equation}
Since 
\[
\overline{\Psi }^{ABCDMN}=-\frac{\sqrt{2}}8N\epsilon ^{MA}{}\psi
{}^{DC}\epsilon {}^{BN}+\frac{\sqrt{2}}8N\psi {}^{DM}\epsilon
{}^{AC}\epsilon {}^{BN}, 
\]
and 
\[
\Theta {}^{ABCDMN}=-\frac{2\sqrt{2}N}3\varphi ^{MD}\epsilon {}^{AC}\epsilon
{}^{BN}-\frac{\sqrt{2}}{24}N\varphi {}^{AD}{}\epsilon ^{MC}\epsilon {}^{BN}-%
\frac{2N^{EM}}3\varphi _E{}^D\epsilon {}^{AC}\epsilon {}^{BN}+\frac 13%
N^{CD}\varphi {}^{AM}\epsilon {}^{BN}, 
\]
the condition 
\[
\Theta ^{ABCDMN}=\overline{\Psi }^{ABCDMN} 
\]
leads to 
\begin{eqnarray*}
&&-\frac{2\sqrt{2}N}3\varphi ^{MD}\epsilon {}^{AC}\epsilon {}^{BN}-\frac{%
\sqrt{2}}{24}N\varphi {}^{AD}{}\epsilon ^{MC}\epsilon {}^{BN}-\frac{2N^{EM}}3%
\varphi _E{}^D\epsilon {}^{AC}\epsilon {}^{BN}+\frac 13N^{CD}\varphi
{}^{AM}\epsilon {}^{BN} \\
&=&-\frac{\sqrt{2}}8N\epsilon ^{MA}{}\psi {}^{DC}\epsilon {}^{BN}+\frac{%
\sqrt{2}}8N\psi {}^{DM}\epsilon {}^{AC}\epsilon {}^{BN},
\end{eqnarray*}
which means 
\begin{equation}
\varphi {}^{AC}{}=3\psi {}^{AC},
\end{equation}
and 
\begin{equation}
2N^A{}_B+N^{CD}\varphi {}^A{}_C\psi _{BD}=0.
\end{equation}
Using (37) and (20) we obtain 
\begin{equation}
\psi _{CA}\psi ^{AB}=\frac 13\epsilon _C{}^B.
\end{equation}
And, (38) leads to trivial result $0=0.$

Similarly, 
\[
\overline{\Pi }{}^{MNCDAB}=-\frac{\sqrt{2}}{24}N\epsilon {}^{MC}\epsilon
^{DA}\epsilon {}^{NB}{}+\frac{\sqrt{2}}{24}N\varphi ^{MA}{}\psi
{}^{CD}\epsilon {}^{NB}{}+\frac{2N^{DA}}3\epsilon {}^{MC}{}\epsilon {}^{NB}, 
\]
and 
\[
\Pi ^{ABCDMN}=\overline{\Pi }^{MNCDAB} 
\]
lead to $\epsilon ^{AM}{}=-\epsilon ^{MA}{},$ and $N^{DC}=N^{DC},$ which are
trivial.

In summary, the conditions for the evolution equations of $\zeta _{aA}$, $%
\widetilde{p}^{aA}$ and $\widetilde{q}_{AB}{}$ together with the assumption
(35) reduce to the reality conditions (36) on the metric and a condition
(39) on the constant spatial vector $\psi ^{AB}$. \ 

In this case the polynomial $Q_p(p,q)$ in the equation (19) has the form 
\begin{eqnarray}
&&Q_p(p,q)  \nonumber \\
&=&-\frac{25N}{64\sigma }\zeta ^{aA}{}{}\widetilde{p}^{BC}\widetilde{p}_{BC}+%
\frac N{64\sigma }\widetilde{p}^{aA}\widetilde{p}_B{}^B-\frac N{64\sigma }%
\widetilde{p}^a{}_B\widetilde{p}^{AB}{}  \nonumber \\
&&+\frac N{64\sigma }\psi {}^{aB}\widetilde{p}_C{}^C\widetilde{q}_B{}^A-%
\frac{3N}{64\sigma }\psi ^{aB}{}\widetilde{p}_B{}^A\widetilde{q}_C{}^C-\frac{%
3N}{32\sigma }\psi ^{aB}{}\widetilde{p}^{AC}\widetilde{q}_{BC}{}{}+\frac N{%
64\sigma }\psi {}^{aB}\widetilde{p}_{BC}{}\widetilde{q}^{AC}  \nonumber \\
&&-\frac N{32\sigma }\psi {}^{aA}\widetilde{p}^{BC}\widetilde{q}{}_{BC}-%
\frac N{64\sigma }\psi {}^{AB}\widetilde{p}_{BC}{}\widetilde{q}^{aC}+\frac N{%
16\sigma }\psi {}_{BC}\widetilde{p}^{CA}\widetilde{q}^{aB}+\frac N{8\sigma }%
\zeta ^{aA}\psi _B{}^C\widetilde{p}^{BD}\widetilde{q}_{CD}{}{}{}  \nonumber
\\
&&-\frac{11N}{192\sigma }\zeta ^{aA}{}\widetilde{q}{}_{BC}\widetilde{q}%
{}^{BC}-\frac N{32\sigma }\widetilde{q}^{aA}\widetilde{q}_B{}^B-\frac N{%
48\sigma }\widetilde{q}^A{}_B\widetilde{q}^{aB}  \nonumber \\
&&-\frac N{16\sigma }\psi ^{aB}\psi {}_C{}^D\widetilde{q}^{CA}\widetilde{q}%
_{BD}-\frac N{64\sigma }\psi ^{aB}{}\psi {}^{CA}\widetilde{q}_C{}^D%
\widetilde{q}{}_{BD}  \nonumber \\
&&+\partial _{(BC)}N^{CB}\widetilde{p}^{aA}-\frac{\sqrt{2}}8\partial
_{(BC)}N\zeta ^{aB}{}\widetilde{p}^{CA}-\frac{\sqrt{2}}8\partial
_{(BC)}N\psi {}^{BC}\widetilde{q}^{aA}-\frac{\sqrt{2}}8\partial _{(BC)}N\psi
{}^{aB}\widetilde{q}^{CA},
\end{eqnarray}
and the evolution equation (23) becomes 
\begin{eqnarray}
&&\stackrel{\cdot }{\widetilde{q}}_{AB}  \nonumber \\
&=&-2\sqrt{2}N\zeta {}^{aD}\psi _D{}^C\partial _{(AC)}{}{}\widetilde{p}_{aB}-%
\frac{\sqrt{2}N}8\zeta {}^{aD}\psi _A{}{}^C{}\partial _{(CD)}\widetilde{p}%
{}_{aB}+2N^{DE}\psi _E{}^C\zeta ^a{}_D\partial _{(AC)}\widetilde{p}%
_{aB}+N^{CE}\psi _{AD}{}\zeta {}{}{}^{aD}\partial _{(CE)}\widetilde{p}{}_{aB}
\nonumber \\
&&-\frac{\sqrt{2}N}{24}{}\partial _{(AC)}\widetilde{q}^C{}_B-\frac{\sqrt{2}N}%
8\psi _{AC}{}\psi {}^{DE}\partial _{(DE)}\widetilde{q}^C{}_B+\frac 23%
N^{CD}\partial _{(AC)}\widetilde{q}_{DB}  \nonumber \\
&&+Q_q(p,q)
\end{eqnarray}
where 
\begin{eqnarray}
&&Q_q(p,q)  \nonumber \\
&=&-\frac{25N}{64\sigma }\psi _{AB}{}{}{}\widetilde{p}^{DC}\widetilde{p}%
_{DC}-\frac{17N}{64\sigma }\psi _{AC}{}\widetilde{p}^{CD}{}\widetilde{p}%
_{BD}{}+\frac{65N}{64\sigma }\psi _{AC}{}\widetilde{p}_B{}{}{}^C\widetilde{p}%
_D{}^D+\psi _D{}^C\widetilde{p}{}_A{}^D\widetilde{p}_{CB}  \nonumber \\
&&+\frac{\sqrt{2}N_{AE}}{8\sigma }\psi _D{}^C\widetilde{p}_{CB}\widetilde{p}%
{}{}^{ED}+\frac{\sqrt{2}N^{DE}}{4\sigma }\psi _A{}^C\widetilde{p}_{CD}{}%
\widetilde{p}_{EB}+\frac{\sqrt{2}N^{DE}}{4\sigma }\psi _D{}^C\widetilde{p}%
_{AB}\widetilde{p}_{CE}{}+\frac{\sqrt{2}N^{DE}}{8\sigma }\psi _D{}^C%
\widetilde{p}_{CB}\widetilde{p}_{AE}  \nonumber \\
&&+\frac N{96\sigma }\epsilon _{AB}\widetilde{p}^{DE}\widetilde{q}{}_{DE}+%
\frac N{64\sigma }{}\widetilde{p}_{AB}{}\widetilde{q}_C{}^C-\frac{35N}{%
32\sigma }\widetilde{p}{}_C{}^C\widetilde{q}_{AB}{}+\frac N{192\sigma }%
\widetilde{p}_A{}{}^C\widetilde{q}_{BC}-\frac{35N}{96\sigma }{}{}\widetilde{p%
}_{BC}\widetilde{q}_A{}^C{}+\frac N{8\sigma }\psi _{AB}{}\psi _E{}^C%
\widetilde{p}^{ED}\widetilde{q}_{CD}  \nonumber \\
&&+\frac N{64\sigma }\psi _{AC}{}\psi _B{}^D\widetilde{p}_{DE}{}\widetilde{q}%
^{CE}-\frac N{16\sigma }\psi _{AC}{}\psi {}_{DE}\widetilde{p}_B{}^E%
\widetilde{q}^{CD}+\frac N{4\sigma }\psi {}^{CD}\psi _A{}^E\widetilde{p}%
{}_{DE}\widetilde{q}_{CB}{}{}-\frac N\sigma \psi {}^{CD}\psi _A{}^E{}{}%
\widetilde{p}_{DB}\widetilde{q}_{CE}  \nonumber \\
&&+\frac{\sqrt{2}N^{DE}}{8\sigma }\psi _{AC}\psi _{FE}\widetilde{p}{}_B{}^F%
\widetilde{q}_D{}{}{}^C-\frac{\sqrt{2}N^{DE}}{2\sigma }\psi _{AC}{}\psi
{}_{FE}\widetilde{p}_B{}^C\widetilde{q}_D{}^F-\frac{\sqrt{2}N^{DE}}{2\sigma }%
\psi _D{}^C\psi _A{}^F\widetilde{p}_{EB}\widetilde{q}_{CF}  \nonumber \\
&&+\frac{\sqrt{2}N^{DE}}{8\sigma }\psi _D{}^C\psi _A{}^F\widetilde{p}_{FB}%
\widetilde{q}_{CE}+\frac{\sqrt{2}N^{DE}}{8\sigma }\psi _F{}^C\psi _{AE}%
\widetilde{p}_D{}{}^F\widetilde{q}_{CB}-\frac{\sqrt{2}N^{DE}}{8\sigma }\psi
_F{}^C\psi {}_{AE}\widetilde{p}_{CB}\widetilde{q}_D{}{}{}^F  \nonumber \\
&&-\frac{11N}{192\sigma }{}\psi {}_{AB}\widetilde{q}{}_{CD}\widetilde{q}%
{}^{CD}+\frac N{16\sigma }\psi {}_{AC}\widetilde{q}{}_B{}^C\widetilde{q}%
_D{}^D+\frac N{48\sigma }\psi {}_{AC}\widetilde{q}{}_{BD}\widetilde{q}^{CD}-%
\frac N{192\sigma }{}\psi _{BC}{}\widetilde{q}{}_{AD}\widetilde{q}{}^{CD}+%
\frac N{48\sigma }\psi {}{}_{CD}\widetilde{q}{}_B{}^C\widetilde{q}_A{}^D 
\nonumber \\
&&+\frac{\sqrt{2}N^{CD}}{6\sigma }\psi _A{}^E\widetilde{q}_{DB}\widetilde{q}%
_{CE}{}-\frac{13\sqrt{2}N^{CD}}{24\sigma }\psi {}_C{}^E\widetilde{q}_{AB}%
\widetilde{q}_{DE}{}{}{}+\frac{\sqrt{2}N^{CD}}{6\sigma }\psi _C{}^E%
\widetilde{q}_{DB}\widetilde{q}_{AE}-\frac{\sqrt{2}N_{AD}}{24\sigma }\psi
_C{}^E\widetilde{q}_{EB}{}\widetilde{q}{}{}{}^{DC}  \nonumber \\
&&+\frac{\sqrt{2}}{24}\partial _{(AC)}N\widetilde{q}^C{}_B-2\sqrt{2}\partial
_{(AC)}N\psi {}^{DC}{}{}\widetilde{p}_{DB}-\frac{\sqrt{2}}8\partial
_{(DE)}N\psi _A{}{}^D{}\widetilde{p}^E{}_B+\frac{\sqrt{2}}8\partial
_{(DE)}N\psi _{AC}{}\psi {}^{DE}\widetilde{q}^C{}_B  \nonumber \\
&&+\frac 23\partial _{(AC)}N^{CD}\widetilde{q}_{DB}-2\partial
_{(AC)}N^{DE}\psi _D{}^C\widetilde{p}_{EB}-\partial _{(DE)}N^{ED}\psi _{AC}{}%
\widetilde{p}^C{}_B.
\end{eqnarray}

\section{Conclusions}

Now we have proved that the evolution equation (17), (19) and (41)
constitute a symmetric hyperbolic system under the conditions (35), (36) and
(39), and the constraint equations are (16) and (25). Here we suppose only
the reality of the shift vector $N^{CD}$ and then obtain naturally the
reality of the spatial metric as one of the conditions of the symmetric
hyperbolic system rather than a independent one as in [23].

This work was supported by the National Science Foundation of China No.
10175032.

\end{document}